\documentclass[12pt,dvipsnames]{article}
\usepackage{times}
\usepackage{geometry}
\usepackage[numbers,sort&compress]{natbib}

\usepackage[flushmargin,norule]{footmisc}
\usepackage[table,dvipsnames]{xcolor}
\RequirePackage{graphicx}
\usepackage[utf8]{inputenc}
\usepackage{enumitem,amssymb}
\usepackage{ragged2e}
\newlist{thematic}{itemize}{8}
\setlist[thematic]{label=$\square$}
\usepackage{pifont}

\usepackage[colorlinks]{hyperref}
\hypersetup{citecolor=blue,linkcolor=blue, urlcolor=blue}
\usepackage{doi}
\usepackage{lineno}
\usepackage[subtle]{savetrees}
\usepackage{tipa}
\newcommand{\citeAstro}[1]{\textbf{\hypersetup{citecolor=red}\cite{#1}}}

\begin{document}
\noindent \begin{large}\textit{Astro2020 APC White Paper}\end{large}\\[0.1cm]
\noindent\begin{large}\textbf{The Southern Wide-Field Gamma-Ray Observatory (SWGO):  A Next-Generation Ground-Based Survey Instrument for VHE Gamma-Ray Astronomy}\end{large} \\


\begin{raggedright}

\noindent \textbf{Project Category:} Ground-Based -- Medium

\noindent \textbf{Thematic Areas:} Particle Astrophysics \& Gravitation; Multi-Messenger Astronomy and Astrophysics; Cosmology and Fundamental Physics; Formation and Evolution of Compact Objects \\[0.2cm]


\end{raggedright}

\pagestyle{empty}

\noindent\textbf{Abstract:}
We describe plans for the development of the Southern Wide-field Gamma-ray Observatory (SWGO), a next-generation instrument with sensitivity to the very-high-energy (VHE) band to be constructed in the Southern Hemisphere.
SWGO will provide wide-field coverage of a large portion of the southern sky, effectively complementing current and future instruments in the global multi-messenger effort to understand extreme astrophysical phenomena throughout the universe.
A detailed description of science topics addressed by SWGO is available in the science case white paper~\cite{2019arXiv190208429A}.
The development of SWGO will draw on extensive experience within the community in designing, constructing, and successfully operating wide-field instruments using observations of extensive air showers.
The detector will consist of a compact inner array of particle detection units surrounded by a sparser outer array.
A key advantage of the design of SWGO is that it can be constructed using current, already proven technology.
We estimate a construction cost of 54M USD and a cost of 7.5M USD for 5 years of operation, with an anticipated US contribution of 20M USD ensuring that the US will be a driving force for the SWGO effort.
The recently formed SWGO collaboration will conduct site selection and detector optimization studies prior to construction, with full operations foreseen to begin in 2026. Throughout this document, references to science white papers submitted to the Astro2020 Decadal Survey with particular relevance to the key science goals of SWGO, which include unveiling Galactic particle accelerators~\citeAstro{2019BAAS...51c.311F,2019BAAS...51c.183D,2019BAAS...51c.513G,2019BAAS...51c.115C,2019BAAS...51c.267H,2019BAAS...51c.122F,2019BAAS...51c.131S,2019astro2020T.459F,2019BAAS...51c.194N}, exploring the dynamic universe~\citeAstro{2019astro2020T.444F,2019BAAS...51c...5J,2019BAAS...51c..92R,2019BAAS...51c.265W,2019BAAS...51c.185V,2019BAAS...51c.553V,2019BAAS...51c.228S,2019BAAS...51c.357S,2019BAAS...51c.260B,2019BAAS...51c.250B,2019BAAS...51c.385R}, and probing physics beyond the Standard Model~\citeAstro{2019BAAS...51c.308V,2019BAAS...51c.202A,2019BAAS...51c.203M,2019BAAS...51c.272H}, are highlighted in red boldface. \\

\noindent \textbf{Corresponding/Lead Author:}
Petra Huentemeyer (Michigan~Technological~University);\\
Contact:~\href{mailto:petra@mtu.edu}{petra@mtu.edu}; +1 (906) 487-1229\\
%
%
%
\noindent\textbf{Co-authors/Proposing Team:} 
P. Abreu\footnotemark[1]\,\footnotemark[2],
A. Albert\footnotemark[3],
R. Alfaro\footnotemark[4],
C. Alvarez\footnotemark[5],
R. Arceo\footnotemark[5],
P. Assis\footnotemark[1]\,\footnotemark[2],
F. Barao\footnotemark[1]\,\footnotemark[6],
J. Bazo\footnotemark[7],
J.~F. Beacom\footnotemark[8],
J. Bellido\footnotemark[9],
S. BenZvi\footnotemark[10],
T. Bretz\footnotemark[11],
C. Brisbois\footnotemark[12],
A.~M. Brown\footnotemark[13],
F. Brun\footnotemark[14],
M. Buscemi\footnotemark[15],
K.~S. Caballero-Mora\footnotemark[16],
P. Camarri\footnotemark[17],
A. Carramiñana\footnotemark[18],
S. Casanova\footnotemark[19],
A. Chiavassa\footnotemark[20],
R. Conceição\footnotemark[1]\,\footnotemark[2],
G. Cotter\footnotemark[21],
P. Cristofari\footnotemark[22],
S. Dasso\footnotemark[23]\,\footnotemark[24],
A. De~Angelis\footnotemark[25]\,\footnotemark[26]\,\footnotemark[27],
M. De~Maria\footnotemark[28],
P. Desiati\footnotemark[29],
G. Di~Sciascio\footnotemark[30],
J.~C. Díaz~Vélez\footnotemark[31],
C. Dib\footnotemark[32],
B. Dingus\footnotemark[3],
D. Dorner\footnotemark[33],
M. Doro\footnotemark[25]\,\footnotemark[27],
C. Duffy\footnotemark[34],
M. DuVernois\footnotemark[29]\,\footnotemark[35],
R. Engel\footnotemark[36],
M. Fernandez~Alonso\footnotemark[37],
H. Fleischhack\footnotemark[12],
P. Fonte\footnotemark[1]\,\footnotemark[38],
N. Fraija\footnotemark[39],
S. Funk\footnotemark[40],
J.~A. García-González\footnotemark[4],
M.~M. González\footnotemark[39],
J.~A. Goodman\footnotemark[41],
T. Greenshaw\footnotemark[42],
J.~P. Harding\footnotemark[3],
A. Haungs\footnotemark[36],
B. Hona\footnotemark[12],
A. Insolia\footnotemark[43],
A. Jardin-Blicq\footnotemark[44],
V. Joshi\footnotemark[40],
K. Kawata\footnotemark[45],
S. Kunwar\footnotemark[44],
G. La~Mura\footnotemark[1],
J. Lapington\footnotemark[34],
J.-P. Lenain\footnotemark[46],
R. López-Coto\footnotemark[27],
K. Malone\footnotemark[3],
J. Martinez-Castro\footnotemark[47],
H. Martínez-Huerta\footnotemark[48],
L. Mendes\footnotemark[1],
E. Moreno\footnotemark[49],
M. Mostaf\'a\footnotemark[37],
K.~C.~Y. Ng\footnotemark[50],
M.~U. Nisa\footnotemark[51],
F. Peron\footnotemark[28],
A. Pichel\footnotemark[23],
M. Pimenta\footnotemark[1]\,\footnotemark[2],
E. Prandini\footnotemark[27]\,\footnotemark[52],
S. Rainò\footnotemark[53],
A. Reisenegger\footnotemark[54],
J. Rodriguez\footnotemark[14],
M. Roth\footnotemark[36],
A. Rovero\footnotemark[23],
E. Ruiz-Velasco\footnotemark[44],
T. Sako\footnotemark[45],
A. Sandoval\footnotemark[4],
M. Santander\footnotemark[55],
K. Satalecka\footnotemark[56],
M. Schneider\footnotemark[41],
H. Schoorlemmer\footnotemark[44],
F. Schüssler\footnotemark[14],
R.~C. Shellard\footnotemark[57],
A. Smith\footnotemark[41],
S. Spencer\footnotemark[21],
W. Springer\footnotemark[58],
P. Surajbali\footnotemark[44],
K. Tollefson\footnotemark[51],
B. Tomé\footnotemark[1],
I. Torres\footnotemark[18],
A. Viana\footnotemark[48],
T. Weisgarber\footnotemark[29]\,\footnotemark[35],
R. Wischnewski\footnotemark[56],
A. Zepeda\footnotemark[59],
B. Zhou\footnotemark[8],
H. Zhou\footnotemark[3]\\

\thispagestyle{empty}
\pagestyle{plain}
\setcounter{page}{0}
\subsection*{\textcolor{BlueGreen}{Introduction}}
\vspace{-.25cm}
This white paper presents our vision and plans for the Southern Wide-field Gamma-ray Observatory (SWGO, \url{www.swgo.org}). SWGO is a next-generation, ground-based survey instrument that will provide a unique view on gamma-ray and cosmic-ray emission from tens of GeV to hundreds of TeV. The facility will improve upon the success of the HAWC Gamma-ray Observatory in Mexico that is surveying the Northern gamma-ray sky with nearly 100\% duty cycle and an instantaneous field of view of $\sim$ 2\,sr. Since 2015, HAWC has discovered new TeV sources and source classes, set new world-leading limits on dark matter decay and annihilation, and played a crucial role in multi-messenger observations \cite{2015ApJ...800...78A,2017ApJ...841..100A,2017A&A...607A.115I,2017ApJ...842...85A,2017ApJ...843...39A,2017ApJ...843...40A,2017ApJ...843...88A,2017ApJ...843..116A,2017ApJ...848L..12A,2017Sci...358..911A,2018ApJ...853..154A,2018JCAP...02..049A,2018JCAP...06..043A,2018Sci...361.1378I,2018PhRvD..98l3012A,2018Natur.562...82A,2019arXiv190512518H}. The success of the water Cherenkov technology implemented by HAWC has inspired an ambitious Chinese-lead effort, LHAASO, in the Northern Hemisphere, which uses a similar design \cite{2019arXiv190502773B}.

Recent years have seen a wealth of paradigm-shifting discoveries including a kilonova associated with merging neutron stars \cite{2017ApJ...848L..12A}, a gamma-ray burst (GRB) with photons detected above 300 GeV \cite{2019ATel12390....1M}, and a detection of a sub-PeV neutrino from a flaring active galactic nucleus (AGN) \cite{2018Sci...361.1378I}. The global multi-messenger astrophysics community recognizes the importance of facilities in both hemispheres that continuously survey the gamma-ray sky in the space and time domain. Wide-field-of-view observatories can not only provide prompt alerts of transient events to the astrophysics community, but they also retain archival information about gamma-ray emission covering large regions in the sky.

Our international collaboration is committed to the ideals of open science. All gamma-ray data will be made publicly available after a brief proprietary period. We envision an approach similar to those practiced by existing NASA missions such as Fermi, Swift, and NuSTAR, or the planned next-generation imaging atmospheric Cherenkov Telescope Array, CTA \cite{2019scta.book.....C}, which provide a data archive and science tools for the astrophysics and astronomy community. In addition, we will explore ways in which communities beyond astrophysics, e.g. cosmology and particle physics, may use data at all levels from the observatory. We propose, for the first time, to provide a formal guest investigator program for a TeV all-sky instrument. While the team members have experience in and are working on a number of methods to make data from their respective observatories public, this represents an even more ambitious and sweeping plan for data dissemination that is not restricted only to high-level data products. The goal is for observatory data to be prepared in a way that will make it straightforward for a diverse science community to access and combine them – a necessity for any astrophysics experiment in the multi-messenger era.

\subsection*{\textcolor{BlueGreen}{Key Science Goals and Objectives}}
\begin{figure}[!ht]
\begin{center}
\includegraphics[width=1\textwidth]{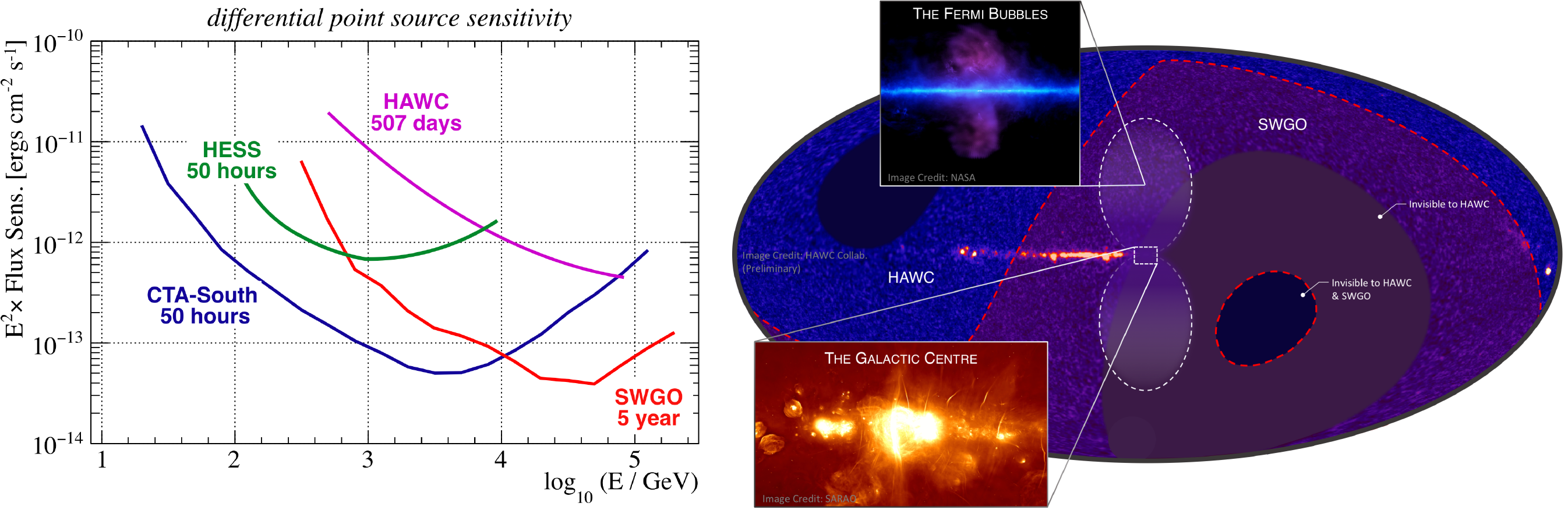}
\caption{
\emph{Left:} Differential point-source sensitivity as a function of energy for the proposed SWGO detector compared to other existing or proposed instruments. \emph{Right:} Sky coverage of SWGO in Galactic coordinates overlaid on HAWC significance map containing over 50 sources. 
}
\label{fig:sensitivity}
\vspace{-0.7cm}
\end{center}
\end{figure}

\vspace{-.25cm}
The key science objectives of SWGO appear in Table~\ref{tab:objectives}, along with the design requirements necessary to achieve them.
The anticipated sensitivity of SWGO is shown in the left panel of Figure~\ref{fig:sensitivity}, in comparison with the existing High Energy Stereoscopic System (HESS) instrument, and CTA-South, which is currently under construction.
SWGO will have a total sky coverage of $\sim$8 sr, as shown in the right panel of Figure~\ref{fig:sensitivity}.
In contrast to the imaging atmospheric Cherenkov Telescope (IACT) sensitivities, which apply to the observation of a single source, the SWGO sensitivity applies to a large number of sources throughout its sky coverage simultaneously.
Figure~\ref{fig:sensitivity} also applies strictly to point sources.
Sources with moderate angular extents will reduce the sensitivity for IACTs 
more severely than for SWGO.\\

\begin{table}
\begin{center}
\begin{tabular}{|r | l|}
\hline
\multicolumn{1}{|c |}{\textbf{Objective}} & \multicolumn{1}{| c|}{\textbf{Major Requirements}} \\
\hline
Measure TeV halos around nearby PWNs & wide field of view \\
Identify sources of PeV Galactic cosmic rays & sensitivity to highest energies \\
Measure the Galactic Center morphology & wide field of view, sensitivity to highest energies \\
Study the nature of the Fermi bubbles & wide field of view \\
Measure the Galactic diffuse emission & wide field of view \\
Measure the local cosmic-ray anisotropy & wide field of view \\
Measure solar cosmic-ray interactions & large unbiased duty cycle \\
Search for new Galactic VHE emitters & wide field of view, large unbiased duty cycle \\
\hline
Detect AGN flares and issue alerts & large unbiased duty cycle, sensitivity $\lesssim$1 TeV \\
Search for periodicity in AGNs & large unbiased duty cycle \\
Measure long-term emission from AGNs & large unbiased duty cycle \\
Search for neutrino VHE counterparts & large unbiased duty cycle \\
Search for counterparts to GW events & wide field of view, large unbiased duty cycle \\
Measure nearby bright GRBs & wide field of view, sensitivity $\lesssim$1 TeV \\
\hline
Search for dark matter annihilation/decay & wide field of view, sensitivity to highest energies \\
Probe Lorentz invariance violation & sensitivity to highest energies \\
Search for PBHs and ALPs & wide field of view \\
\hline
\end{tabular}
\end{center}
\caption{Key objectives for SWGO and the most important requirements for achieving them.}\label{tab:objectives}
\end{table}

\noindent \textbf{Unveiling Galactic particle accelerators --}
Sensitivity to astrophysical particle accelerators in the local Galactic neighborhood is one of the greatest strengths of SWGO.
The detection of Geminga and PSR B0656+14 with HAWC suggests that nearby pulsars 
strongly influence their surroundings~\cite{2017Sci...358..911A}, leading to a new understanding of particle propagation in the vicinity of PWNs
~\cite{2017PhRvD..96j3016L}.
The large angular extents of these TeV halos allow us to study the propagation of particles within them in unprecedented detail~\citeAstro{2019BAAS...51c.311F}.
Locating SWGO in the southern hemisphere will place within reach $\sim$12 Geminga-like middle-aged pulsars within 1 kpc of Earth for study, along with many older and/or more distant pulsars likely to have similar TeV halos.
A thorough understanding of these local accelerators is necessary for interpreting the unexpected excess of positrons observed at Earth~\cite{2009Natur.458..607A}, either as being due to a local source or some more exotic mechanism~\citeAstro{2019BAAS...51c.183D,2019BAAS...51c.513G}.

Sources of cosmic rays with energies in excess of 1 PeV are expected to produce VHE gamma rays above 100 TeV~\citeAstro{2019BAAS...51c.115C,2019BAAS...51c.267H}.
With its sensitivity to the highest energies, SWGO is uniquely suited to detect these sources and measure any cutoffs in their spectra that indicate the maximum energy to which they are able to accelerate particles.
Strong evidence exists that such a source exists in the Galactic Center~\cite{2016Natur.531..476H}.
Measurements of the Galactic Center will take advantage of SWGO's wide field of view to provide energy-dependent morphology measurements for characterizing the propagation of particles in its vicinity.
A combined study with VHE gamma rays and neutrinos also has the potential to reveal the processes responsible for these high-energy particles.

SWGO will play an important role in exploring the nature of extended structures, including the Fermi bubbles~\cite{2014ApJ...793...64A,2017ApJ...842...85A,2019PhRvD..99h3007Y}.
SWGO measurements of the Galactic diffuse emission will reveal spatial differences, like those seen at lower energies~\cite{2016ApJS..223...26A} that are important for understanding the distribution of particle accelerators throughout the Galaxy and the details of how they accelerate cosmic rays~\citeAstro{2019BAAS...51c.122F}.
The accelerated particles themselves will also be accessible to SWGO via precise measurements of Galactic cosmic rays in the PeV region~\citeAstro{2019BAAS...51c.131S}.
The combination of HAWC and/or LHAASO data with cosmic-ray observations from SWGO will permit a nearly all-sky observation of the cosmic-ray anisotropy~\citeAstro{2019astro2020T.459F}, connecting the propagation of cosmic rays to the structure of the local heliospheric and interstellar magnetic fields~\cite{2019ApJ...871...96A}.

In the immediate solar neighborhood, SWGO will play a primary role in investigating the production of VHE gamma rays produced in cosmic-ray interactions with the solar atmosphere~\citeAstro{2019BAAS...51c.194N}, since few existing and planned TeV gamma-ray detectors can monitor the Sun.
Observations of the Sun by SWGO will constrain the production mechanism of these gamma rays and its relationship to the solar cycle, which are currently unknown~\cite{1991ApJ...382..652S,2011ApJ...734..116A,2016PhRvD..94b3004N,2017PhRvD..96b3015Z,2018PhRvL.121m1103L,2018PhRvD..98f3019T}.
Combining the observations with searches for solar atmospheric neutrinos will provide even better constraints~\cite{1996PhRvD..54.4385I,2017PhRvD..96j3006N,2017JCAP...07..024A,2017JCAP...06..033E,2017ICRC...35..965I}.
Finally, measurements of the ``shadow'' in the flux of Galactic cosmic rays created by the Sun will provide independent probe of the magnetic fields near the Sun and in the interplanetary environment~\cite{2013PhRvL.111a1101A,2018PhRvL.120c1101A,2019arXiv190312638B}.
We also anticipate unexpected detections similar to the recent HAWC detection of VHE emission coincident with the lobes of the compact binary SS 433~\cite{2018Natur.562...82A}.
This unexpected result, achieved due to the excellent sensitivity of HAWC above 10 TeV, revealed leptonic particle acceleration in astrophysical jets for the first time. The Galaxy may well harbor several other VHE sources with spectra peaking at these energies; a surface array in the southern hemisphere with even greater sensitivity will be uniquely suited to observe them.
If the sources are transient, then SWGO's large unbiased duty cycle, in addition to its wide field of view, will be important for these efforts.\\

\noindent \textbf{Exploring the dynamic universe --}
Blazars---active galactic nuclei (AGNs) with a relativistic jet oriented along the line of sight to Earth---dominate the extragalactic VHE sky.
Because blazars are highly variable, the monitoring capabilities of SWGO will be ideal for detecting strong flares, such as that from PKS 2155-304 in 2006~\cite{2007ApJ...664L..71A}, which can constrain the VHE production mechanisms of these highly luminous, extreme objects~\citeAstro{2019astro2020T.444F,2019BAAS...51c...5J,2019BAAS...51c..92R,2019BAAS...51c.265W}.
Monitoring at a regular cadence will also enable efforts to detect periodic components unambiguously in select VHE blazar spectra, for which evidence has been presented~\cite{2006APh....26..209O, 2017ApJS..232....7F} and which may be indicative of binary black hole systems at the center of some active galaxies~\cite{2003ASPC..299...83R}.
The past decade has seen the emergence of a population of extreme blazars with relatively little variability~\cite{2015MNRAS.451..611B} whose stability may result from cosmic-ray interactions along the line of sight~\cite{2011ApJ...733L..21D,2012ApJ...749...63M}.
Monitoring extreme blazars on years-long time scales would help to test these models, and it would provide important information for constraints on the intergalactic magnetic field~\cite{2018ApJS..237...32A}.
The large unbiased duty cycle of SWGO is crucial for monitoring AGNs in general and blazars in particular, and the wide field of view will enable many AGNs not yet detected in the VHE band to be monitored as well.
Sensitivity to gamma rays below 1 TeV is essential due to attenuation on the extragalactic background light \cite{1966PhRvL..16..252G}.
Plans for achieving the requisite low-energy sensitivity as shown in Table~\ref{tab:obs} are outlined in the Technical Overview section of this paper.

AGNs are favored candidates for the acceleration sites of the highest energy cosmic rays, which are broadly understood to be extragalactic in origin \cite{2007Sci...318..938P, 2016ApJ...830...81F}.
Recent support for this hypothesis has come from the IceCube detection of a high-energy neutrino in coincidence with a VHE flare from the blazar TXS 0506+056~\cite{2018Sci...361.1378I}, with additional support from a search of archival IceCube data that revealed a neutrino flare~\cite{2018Sci...361..147I}.
Due to its all-sky monitoring capability, HAWC was able to follow up the results of this search and place a constraining limit on the VHE gamma-ray emission during this flare (paper in preparation), which is not possible for the IACTs.
The unbiased monitoring capabilities of SWGO will play an important role in searching for the VHE counterparts of neutrinos detected by the upcoming advanced instruments IceCube-Gen2 and Km3NeT~\citeAstro{2019astro2020T.444F,2019BAAS...51c.228S,2019BAAS...51c.553V,2019BAAS...51c.185V}.

The association of GRB 170817A with the detection of a compact merger event in gravitational waves~\cite{2017ApJ...848L..12A}, coupled with the discovery that GRB 190114C spectra can extend to energies beyond 300 GeV~\cite{2019ATel12390....1M}, has begun to illuminate the nature of these brief but highly luminous events.
Due to the limited fields of view and typical duty cycles of $\sim$15\% for the IACTs, there is a pressing need for a wide-field instrument with sufficient sensitivity in the 0.1--1 TeV energy range to monitor the sky for both isolated GRBs and those coincident with GW detections, ultimately leading to a population of VHE-detected GRBs for understanding their production mechanisms~\citeAstro{2019BAAS...51c.228S,2019BAAS...51c.553V,2019BAAS...51c.357S,2019BAAS...51c.260B,2019BAAS...51c.250B,2019BAAS...51c.385R}.
Searches for FRBs~\cite{2016PASA...33...45P} or other unexpected VHE transient phenomena will also rely on the unique monitoring capabilities of SWGO.
Sub-threshold events that are not significant on their own will be made available for joint analysis via the Astrophysical Multi-messenger Observatory Network~\cite{2013APh....45...56S}, where their combination with data from other instruments is likely to lead to new discoveries.\\

\noindent \textbf{Probing physics beyond the Standard Model --}
SWGO will scrutinize nearby extended sources of dark matter for evidence of gamma rays produced in annihilation or decay processes.
Weakly interacting massive particles (WIMPs) that were once in thermal equilibrium in the early universe remain one of the most promising explanations for dark matter.
Thermal WIMPs with masses between $\sim$2 TeV and $\sim$100 TeV are largely still viable~\cite{2019arXiv190411503S}, and
only astrophysical experiments similar to SWGO can probe heavy dark matter ($>$1 TeV).
CTA will probe thermal WIMPs from 100 GeV to 10 TeV, leaving the heaviest of thermal WIMP masses unconstrained.
SWGO, along with current and future gamma-ray experiments, will probe nearly the full thermal WIMP mass range by extending the sensitivity up to 100 TeV.
Also, both SWGO and CTA will be sensitive to thermal WIMPs for masses from about 1 to 10 TeV.
If there is a gamma-ray signal in that mass range, both experiments should see it, leading to independent confirmation of the signal. 

SWGO searches for gamma rays from dark matter interactions will be most sensitive in the inner Galactic halo~\cite{2019arXiv190603353V}.
Widely varying models for the distribution of dark matter in the Galactic halo pose little challenge for an instrument that routinely monitors a large fraction of the entire sky~\citeAstro{2019BAAS...51c.308V}.
Due to its wide field of view and southern latitude, SWGO will have no need to optimize its observing strategy for different dark matter profiles.
Recent deep surveys have identified numerous new ultra-faint dwarf spheroidal galaxies, mostly in the southern hemisphere (see e.g.~\cite{2015ApJ...807...50B}), that will serve as promising targets for the production of VHE gamma rays by dark matter~\citeAstro{2019BAAS...51c.202A,2019BAAS...51c.203M}.
SWGO will also test models in which dark matter trapped in the sun annihilates into a metastable mediator that decays to gamma rays outside of the sun~\cite{2010PhRvD..81g5004B,2010PhRvD..82k5012S,2011PhRvD..84c2007A,2017PhRvD..95l3016L,2018PhRvD..98l3012A,2019arXiv190311363N}.


In addition to dark matter studies, searches for unconventional physics such as primordial black holes (PBHs) or axion-like particles (ALPs) will leverage the high-energy sensitivity and all-sky monitoring of SWGO to place strong constraints on these theories~\citeAstro{2019BAAS...51c.272H}.
Lorentz invariance violation, which can cause the highest energy photons to undergo decay~\cite{2017PhRvD..95f3001M}, will also be tested~\citeAstro{2019BAAS...51c.203M}.


\subsection*{\textcolor{BlueGreen}{Technical Overview}}
\vspace{-.25cm}


When a high-energy gamma ray enters the atmosphere, it initiates the production of a particle cascade commonly known as an extensive air shower (EAS). The particles in these air showers can be recorded with arrays of particle detectors. Charged cosmic nuclei, commonly known as cosmic rays, also initiate EASs and are the dominant background contribution for ground-based gamma-ray astronomy. 

For a gamma ray of given direction and energy, the amount of energy recorded by the particle detector array provides the amount of information available for reconstruction and is therefore a key design parameter in the performance optimization of the observatory. The recorded energy scales on average with the following observatory parameters: \emph{elevation}, \emph{particle detection threshold}, \emph{detector ground coverage}, and \emph{instrumented area}. Another key parameter is the ability to discriminate between air showers initiated by gamma rays and cosmic rays.  Gamma rays produce \emph{electromagnetic cascades}, mainly consisting out of electrons (and positrons) and gamma rays produced by pair-production and bremsstrahlung processes respectively. Cosmic rays initiate \emph{hadronic cascades}. As a consequence of the hadronic interactions, sub-showers and muons are produced. These lead to features that can be exploited to discriminate them from pure electromagnetic cascades.

The observatory will continuously record EASs coming from all directions, leading to an instantaneous field of view of roughly 90$^\circ$ ($\pm45^\circ$ from zenith). The amplitude and time of the signals recorded by the individual detection units are combined and used for reconstruction and cascade classification. By projecting the arrival direction of the gamma-ray candidate events onto the celestial sphere, 2/3 of the sky is observed on a daily basis and a deep observation is accumulated during the lifetime of the observatory. The main characteristics of the observatory, estimated by an extrapolation of the HAWC performance \cite{2017ApJ...843...39A}, can be found in Table~\ref{tab:obs}.\\

\begin{table}
\begin{center}


\begin{tabular}{| l | c | c | c | }
\hline
\multicolumn{1}{|c |}{\textbf{Specification}} & \multicolumn{3}{ c|}{\textbf{Performance} } \\
 & 0.3\,TeV & 3\,TeV & 30\,TeV \\
\hline
Angular resolution & 1$^\circ$ & 0.3$^\circ$ & 0.15$^\circ$ \\
Energy resolution & 100\% & 50\% & 25\% \\
Background rejection & 50\% &  99\% & 99.9\%\\
Effective Area & 35 000\,m$^2$ &   221 000\,m$^2$ & 221 000\,m$^2$ \\
\hline
Field of View & \multicolumn{3}{| c|}{90$^\circ$}\\
Pointing accuracy & \multicolumn{3}{ c|}{$< 0.05^\circ$}\\
\hline
\end{tabular}

\end{center}
\caption{Observatory specifications as function of gamma-ray energy. The numbers presented correspond to vertical incoming gamma rays for an observatory at 5\,km altitude.}\label{tab:obs}
\end{table}

\begin{figure}[!ht]
\begin{center}
\includegraphics[width=0.45\textwidth]{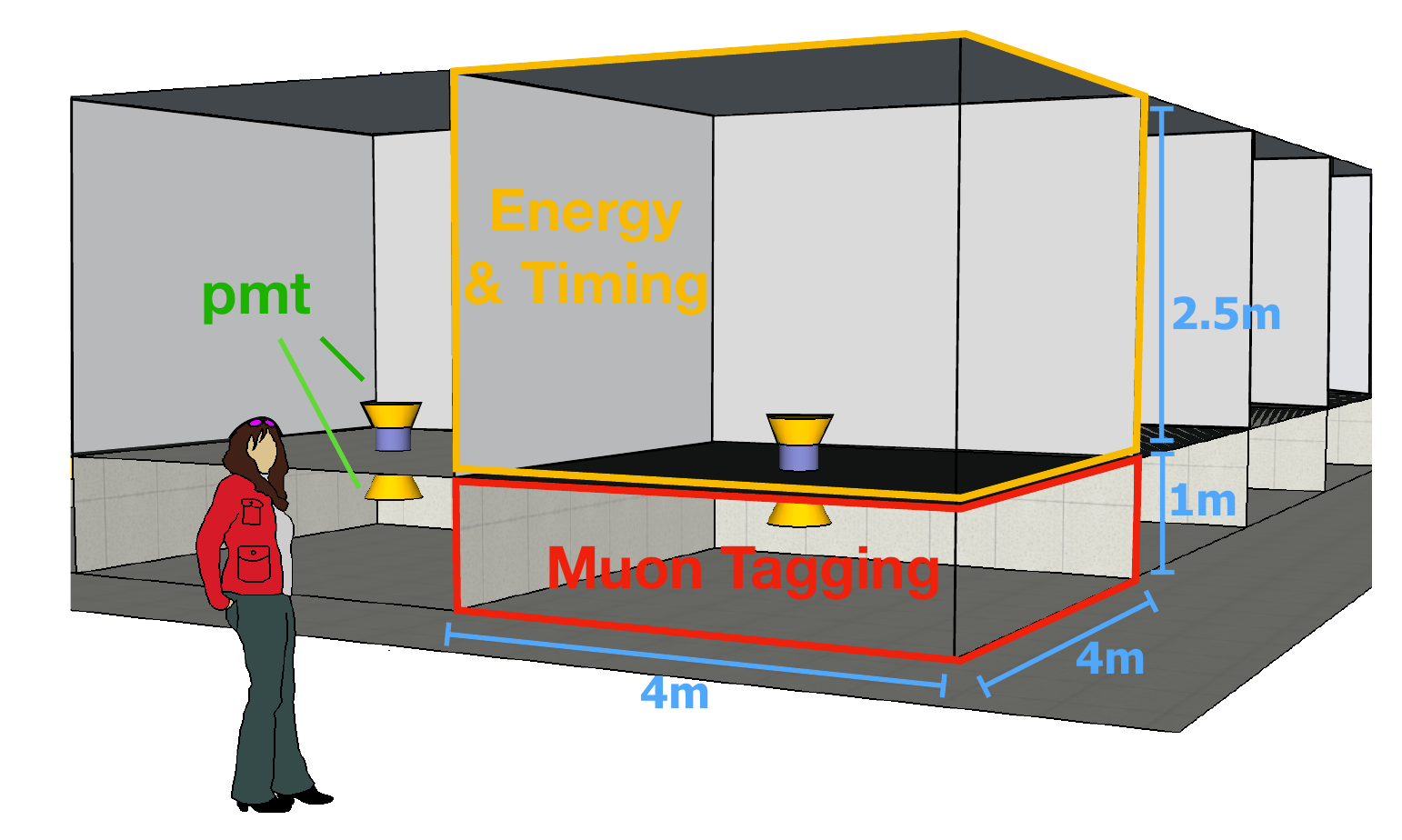}
\includegraphics[width=0.45\textwidth]{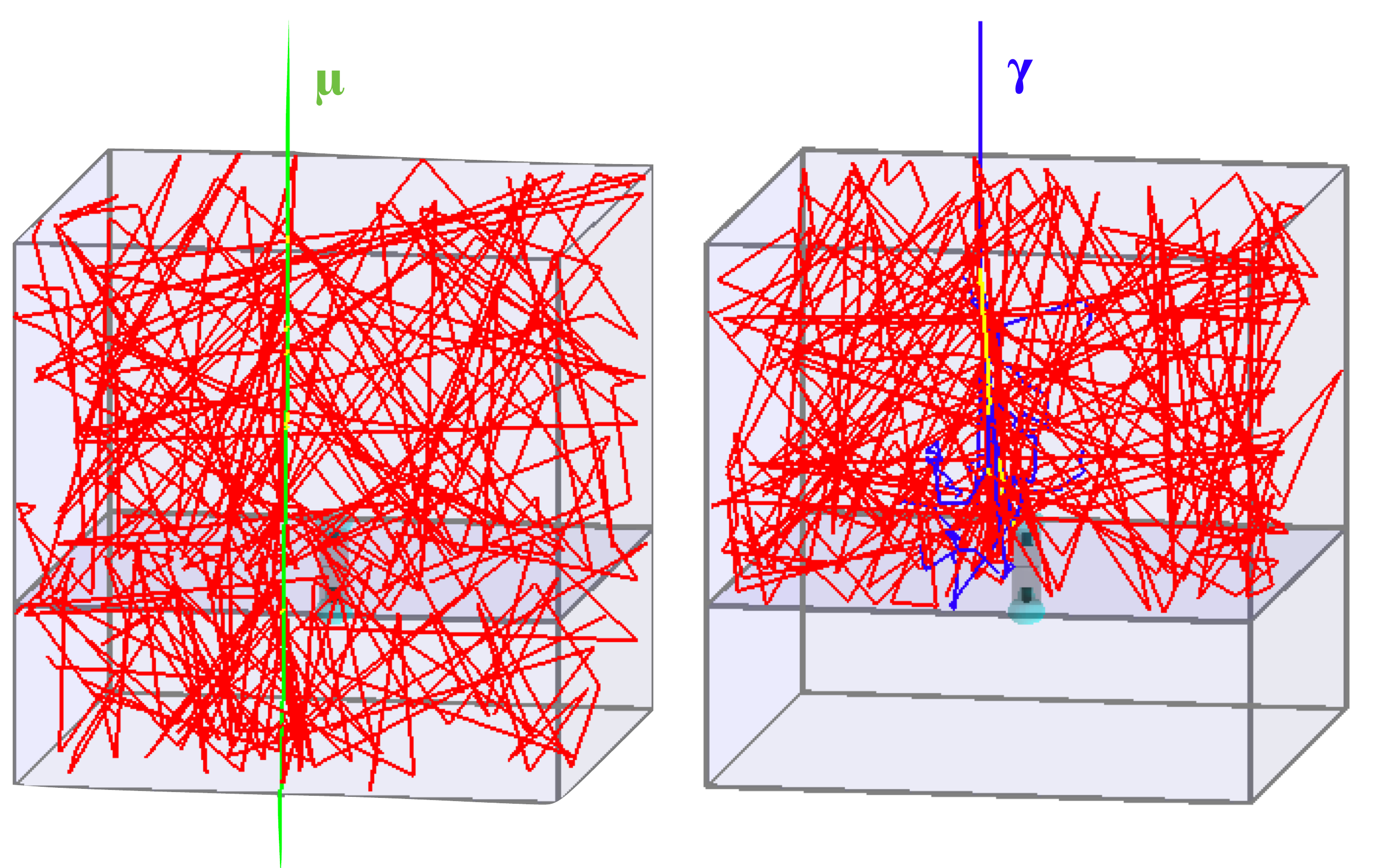}
\caption{\emph{Left: }Illustration of detection unit design. \emph{Right: }Muon and gamma-ray passing through a detection unit. Red lines indicate a fraction of the Cherekov photon tracks, while the green, blue, and yellow lines indicate the tracks of muons, gamma rays and electrons respectively.}
\label{fig:unit}
\vspace{-0.7cm}
\end{center}
\end{figure}
\begin{figure}[!ht]
\begin{center}
\includegraphics[width=0.4\textwidth]{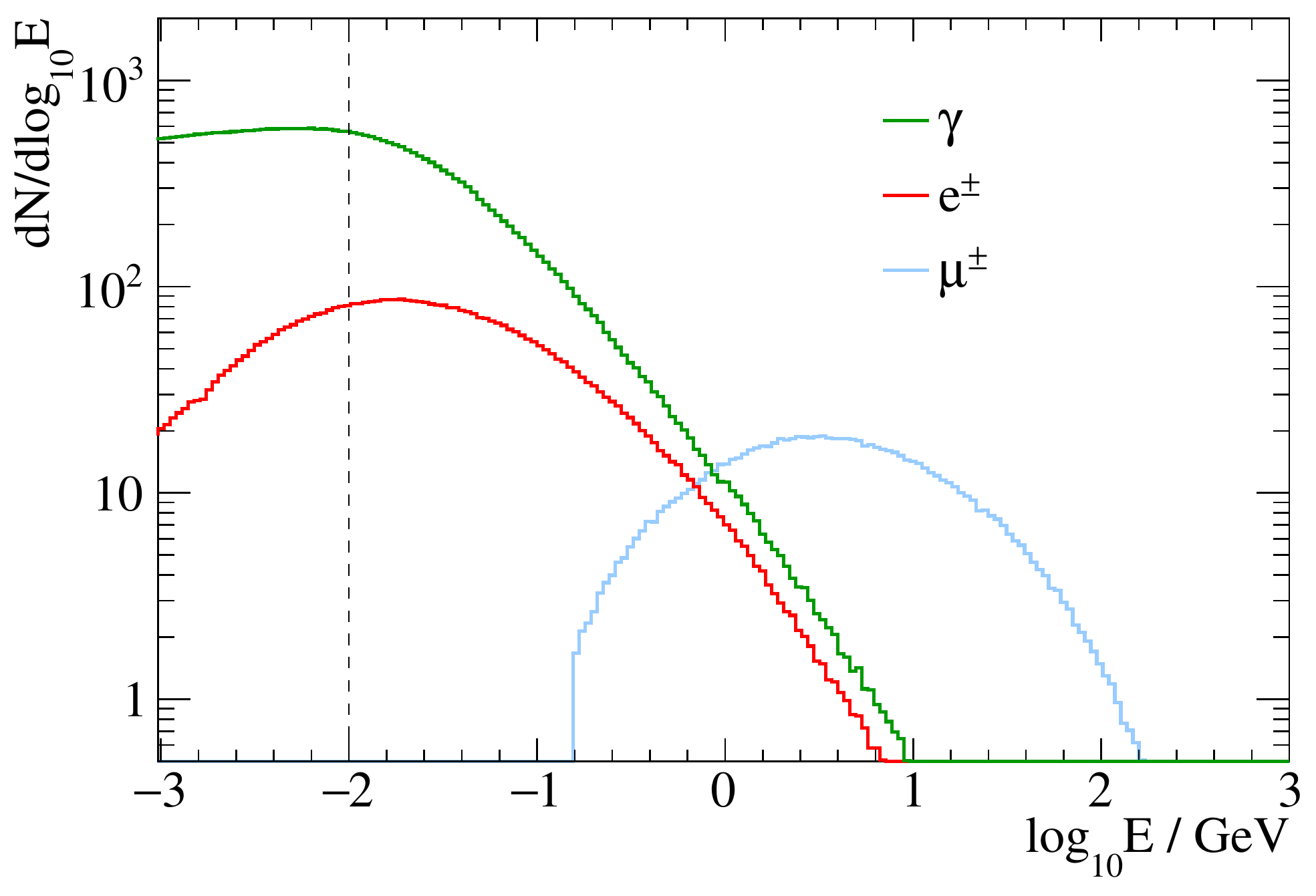}
\includegraphics[width=0.4\textwidth]{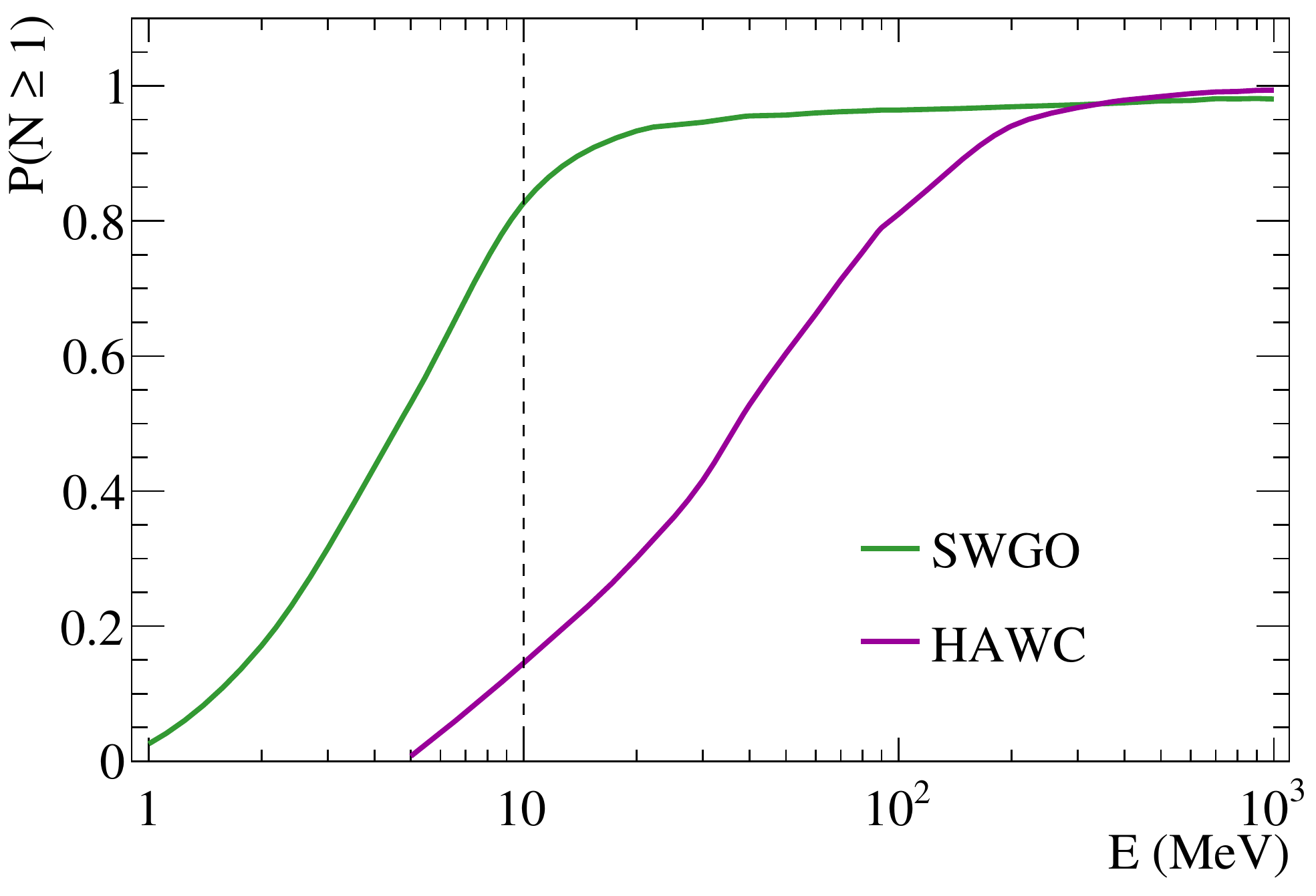}
\caption{\emph{Left:} Average number density at 5\,km above sea level for air showers initiated by 1 TeV protons (adapted from \cite{Schoorlemmer2019}). \emph{Right: } Detection threshold probability for secondary gamma rays of the SWGO unit design. The top compartment is compared to a HAWC detector unit. A 8" PMT R5912 Hamamatsu PMT has been assumed in this simulation. The vertical dashed line in both panels indicates 10\,MeV.}
\label{fig:UnitThres}
\vspace{-0.7cm}
\end{center}
\end{figure}

\noindent \textbf{Particle detection units, array layout, read-out, and triggering --}
The core functionality of a single particle detection unit is to provide accurate local measurements of the arrival time and energy density of the particles in air showers.  We proposed a double-layer water Cherenkov detection unit design as shown in Figure~\ref{fig:unit}.  A maximally densely packed array will be constructed from optically isolated units. Depending on the available infrastructure and the layout of the selected site, these units will either be constructed by segmenting a large volume of water (like in LHAASO), or will consist of independent ``tanks'' (like HAWC).
%
%
The proposed detection unit design is an evolution of the HAWC design, made possible by improvements in technology and cost-reduction of electronics over the last decade. This makes it feasible for SWGO to detect both direct and reflected Cherenkov photons, and to employ and read out more photon sensors, allowing for smaller detection units than HAWC. This results in a significant improvement in sensitivity in the sub-TeV energy domain, as seen in Figure 3.
The science objectives require a significantly improved sensitivity in the sub-TeV energy domain. For a detection unit this means lowering the energy threshold to individual particles, which will be achieved by the top chamber of the proposed detection unit. The walls of the water volume will be lined with white diffusive reflective material, like Tyvek as used in the Pierre Auger Observatory \cite{2008NIMPA.586..409P}. This will significantly enhance the number of detectable particles per air shower and therefore both aid the triggering and reconstruction of low-energy showers.  On the floor of the upper compartment a large photomultiplier-tube (PMT) of $8"$ diameter is installed to measure the Cherenkov light. The PMT faces upwards in order to be able to measure the arrival time of direct Cherenkov light, which is crucial for accurate timing to be used for direction reconstruction.
As can been seen from the distributions shown in Figure~\ref{fig:UnitThres}, the most common particles in air showers are gamma rays in the $\sim$10\,MeV domain. Here we picked the average distributions for 1 TeV protons, but the shapes of the gamma-ray and electron distributions are roughly independent of primary particle type, energy, and detector elevation. The detection probability, estimated by a full GEANT4 simulation, shows that the top chamber of the proposed design has more than double the detection probability for 10\,MeV gamma rays compared to HAWC.

The bottom chamber of the detection unit has a two-fold purpose: muon identification and increasing the dynamic range. Muons are much more prevalent in hadronic cascades and the unambiguous identification of muons in an event will aid in background rejection over the full energy range of the observatory. The typical muon energy in air showers is on the order of a few GeV (see Figure~\ref{fig:UnitThres}). At these energies muons travel straight through both top and bottom chambers, depositing a significant amount of Cherenkov light in both. High-energy gamma rays and and electrons will cascade in the top chamber and only a small fraction (if any) of this cascade can make it to the bottom chamber. Hence, the light ratio between the top and bottom chamber can be used to identify muons. Additionally, the bottom chamber provides a large dynamic range in the shower core, where for the highest energy gamma rays the particle density is very high. The PMT in the top compartment will saturate, but only a few particles will penetrate to the bottom compartment which can still be used to provide a good estimate of the particle density. This will aid the high-energy performance of the observatory. As in the top compartment, the walls are reflective; however, the PMT faces downwards as timing accuracy is not crucial. In this configuration, the top and bottom PMTs can be fitted in a single module which will ease maintenance and installation. The size of the PMT in the bottom compartment can be smaller and preliminary results indicate that a $3"$ diameter PMT should be sufficient to perform muon identification.

\begin{figure}[!ht]
\begin{center}
\includegraphics[width=.9\textwidth]{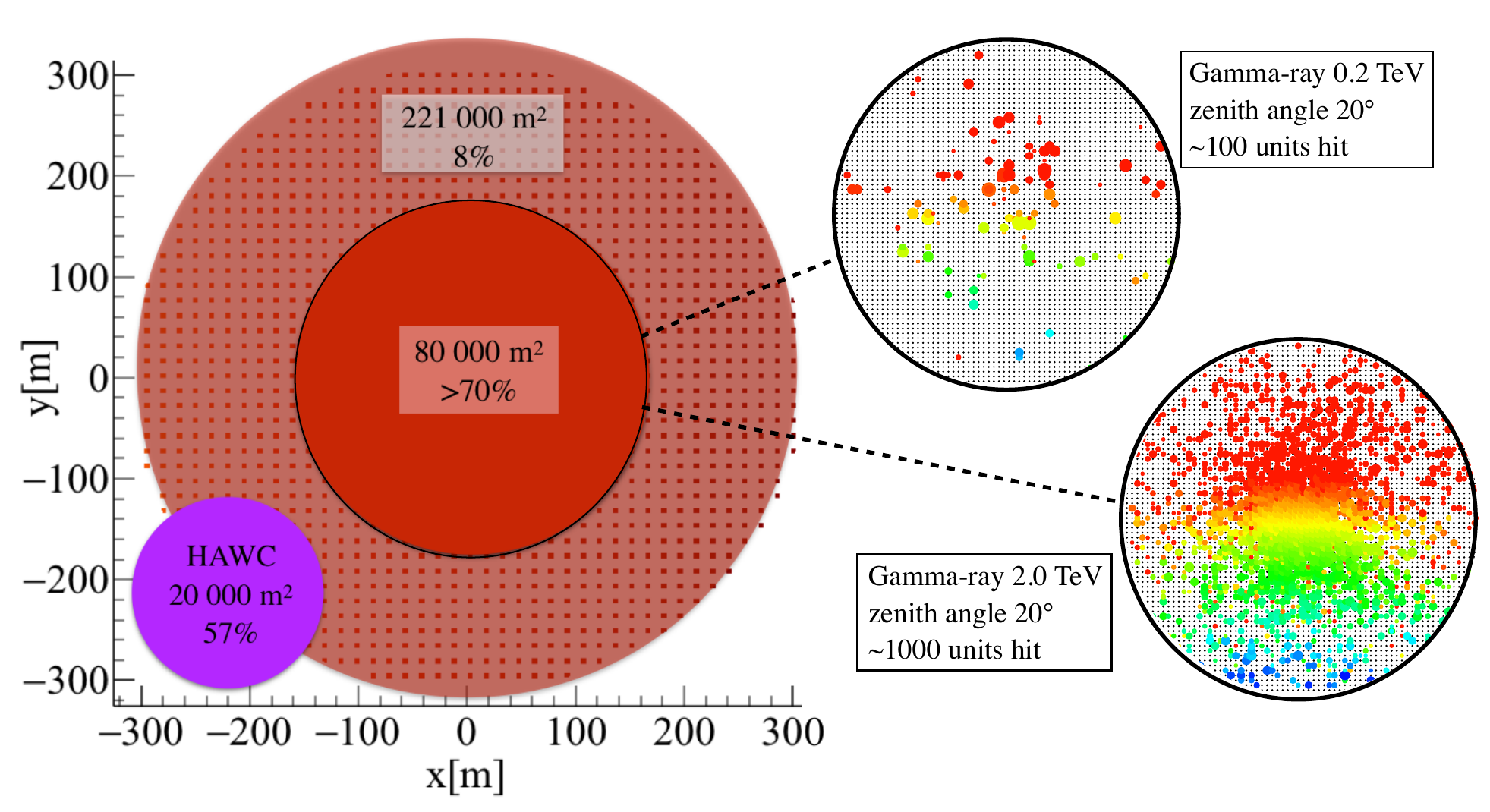}
\caption{\emph{Left: }Array configuration with instrumented areas and ground coverage indicated (and HAWC equivalent for comparison).  \emph{Right: } SWGO response for two simulated gamma-ray events. The color coding indicates the time gradient. The 200\,GeV example illustrates the large particle spread even at low energy.}
\vspace{-0.7cm}
\label{fig:layout}
\end{center}
\end{figure}

The array is foreseen to consist of a densely populated inner part with 4000 detection units, surrounded by a sparser outer array of 1000 detection units as seen in a sketch of a possible layout in Figure~\ref{fig:layout}. The inner part will drive the low-energy performance of the observatory and the outer part will extend the effective area to improve the high-energy reach. 
The dense inner array will cover roughly four times the area of the HAWC Observatory's main array, with higher ground coverage. Besides the obvious increase in flux sensitivity due to the larger area, the significant increase in size will have additional advantages. Firstly, the typical spread of particles for air showers below 200\,GeV is $\gg$75\,m ($\gg$150\,m) for gamma rays (protons) (see \cite{Schoorlemmer2019} for more details). Figure~\ref{fig:layout} illustrates the low particle density by showing the response of a simulated detector to a 200\,GeV gamma ray. Therefore, to effectively measure gamma rays below 1 TeV, a dense, large instrumented area is of paramount importance. Secondly, muons typically have significant transverse momentum due to which they arrive at a larger distance from the shower core. Increasing the muon-sensitive area will improve the identification of hadronic particle cascades and reduce the background rate over the full energy range of the observatory.

The sparser outer array is a cost efficient way of boosting the effective area for the highest energy gamma rays. Particle densities for high energy showers landing in the outer array are high enough that basic shower parameters can be accurately reconstructed with fewer detectors. The size of this outer array is such that the observatory will have the highest point source sensitivity in the Southern hemisphere above tens of TeV. 

As in the HAWC observatory, the full array will have a continuous read-out with a global software trigger that is implemented to reduce the data volume to be archived. Several digitization solutions have been developed that can be adapted to match the requirements (single photon-electron threshold and nanosecond timing accuracy). In the R\&D phase the options will be investigated (see Section \emph{Technology Drivers}). The software trigger will run on a computing farm, where real-time reconstruction will also be performed. In case of astrophysical transient events, the multi-messenger community can be alerted on the timescale of a few seconds to a minute.\\

\noindent \textbf{Site and infrastructure --}
The performance needed for a wide field of view gamma-ray survey instrument that accomplishes the science goals presented above places several constraints on the site selection~\cite{2019arXiv190208429A}. The main ones are: latitude between 10 and 40 degrees south, altitude higher than 4500 m, average yearly temperature above freezing, and access to water sources. Existing infrastructure including roads, power, optical cable, etc. is also highly desirable. Several potentially suitable sites between 4800 and 5000\,m in altitude have been identified. Among these are the Atacama Large Millimeter/submillimeter Array (ALMA) site in the high plateau of the Atacama Desert in Chile; the Cerro Vecar in Argentina where QUBIC and LLAMA are being built; the ALPACA site in Bolivia; and Laguna Sibinacocha, the highest altitude large lake, located in Peru.

The observatory requires a flat area of 80,000\,m$^2$ for the main compact array. For the sparse array an area of 220,000\,m$^2$ is needed, which does not need to be flat but is required to be accessible to install individual detection units.  
The average power needs for the observatory will be on the order of 50 KVA. Reliable lower bandwidth internet access is needed for external monitoring and control. The data rate will be on the order of 100 MBytes/s with the data stored on removable drives that will be transported to the nearest site with high enough bandwidth to transfer it to data centers worldwide.

\subsection*{\textcolor{BlueGreen}{Technology Drivers}}
\vspace{-.25cm}
A strong point of this observatory is that it can be completely built with existing technology. However, in order to improve signal quality and reduce use of expensive signal and high-voltage cables and connectors, in the R\&D phase the option of digital-optical-modules to be deployed in each detection unit will be investigated. These modules will host the two photo-multipliers, a high-voltage unit, a field-programmable-gate-array (for control and signal processing), and an analog-to-digital converter. Each module will be connected to a power-cable and two optical fibers, one for the data-output and control, and another providing a clock signal for time synchronization. For all components, except the time synchronization, commercial cost-effective solutions are available. The timing synchronization system needs to be developed, but can be built upon the well-established white rabbit protocol \cite{WhiteRabbit}. 

\subsection*{\textcolor{BlueGreen}{Organization, Partnerships, and Current Status}}
\vspace{-.25cm}
By signing a statement of interest (SoI) on July 1st 2019, 41 institutions from nine countries have formed a collaboration with the intent to pursue research and development towards SWGO (see \href{https://www.swgo.org/SWGOWiki/doku.php?id=collaboration}{www.swgo.org}  for more information). 
This new collaboration is a step forward beyond the existing SGSO alliance, LATTES and STACEX projects, and the initial duration of the R\&D program is agreed to be three years.
The SoI foresees that the execution of the R\&D program will be overseen by a Steering Committee (SC) consisting of national representatives of the signees. 
The SC will elect a Chair, who acts as spokesperson for the program. 
The following working groups will be founded, with coordinators and deputies appointed by the SC: 1. Science case development, 2. Simulations, analysis and array optimization, 3. Candidate site evaluation, 4. Detector design and development. The working group activities are closely interlinked and coordination between them is essential. 
Individual scientists not affiliated with a member institution of the SWGO collaboration can on request become Supporting Scientists, welcome to join any of the working groups, with full access to program-internal information.  Currently, 11 supporting scientists from 7 countries have signed up. 
The institutional and unaffiliated members of the SWGO collaboration agree to contribute to the R\&D program, using their own resources, via the above working groups. They will offer mutual support towards realization of a detailed proposal for the construction and operation of a southern observatory. 


\subsection*{\textcolor{BlueGreen}{Schedule}}
\vspace{-.25cm}
\textbf{2019-2023:} The site will be selected and final parts designed and prototyped. A small array of detection units will be operated at the site.

\noindent {\textbf{2023-2026:} The site will be leveled and an operations building constructed during the first year. Detection units and the data acquisition building will be installed in the following 2 years.  Operations will begin with a subset of the full array during the second year.}

\noindent {\textbf {2026-2031:} Operations of the full array with 5000 detection units.} 

\noindent {\textbf {2031-2036:} Extended operations period if warranted as determined by external peer review.} 

\noindent {\textbf {2037:} Decomissioning will occur.}

\subsection*{\textcolor{BlueGreen}{Cost Estimate}}
\vspace{-.25cm}
The total construction cost is estimated to be 54M USD as shown in Table 3.  This cost estimate was performed by HAWC collaborators from the Univ. of Maryland, Los Alamos National Lab, and Max Planck Institute in consultation with Argentinian scientists.  Many aspects of the detector design are very similar to HAWC.  All costs are in current US dollars. The highest cost item is for the structural detection units which hold the water and keep out all light.  A currently valid quotation from Aquamate Engineered Water Tanks which includes shipping to the nearest port is the primary basis of the 25M USD estimate.  The next largest cost is 10M USD for PMTs and their associated hardware. This cost is primarily based on procurements within the last year from Hamamatsu.  5M USD is budgeted for the electronics which is based on the HAWC upgrade completed within the last year.  These electronics were custom designed by Max Planck Institute and commercially built.  Costs for water include 0.5M USD for a filtration plant at or near the site and 2.5M USD for procurement and delivery. This is based on the HAWC experience of trucking water from a well 10 miles away, but will vary depending on the site chosen. The site infrastructure costs will also depend on the site chosen, but the cost given of 5M USD is consistent with HAWC experience and Argentinian costs. Laborers will be hired to do the construction and again HAWC experience of the person-years required is used and an average rate of 15 USD/hour. Management and engineering costs as estimated from HAWC experience are 1M USD/year for 3 years of construction.  No contingency is included; however, if costs increase, the number of detection units can be reduced. Also, importation costs are not included.  Negotiating waivers of these duties will be part of the site selection.

The operations costs will be minimal at 1.5M USD/year because the detectors, as demonstrated by HAWC, are very robust with no moving parts. 300k USD/year for infrastructure is required for items such as road and site maintenance, electrical power, internet, and expendables such as batteries for Uninterruptible Power Supplies.  Due to the quantity of data generated, a large number of computers and disk arrays is required both at the site and at data centers at universities, and 800k USD/year is budgeted.  Management, engineering, and technical support is required at 400k USD/year.

Science costs are not included as part of the project as is customary for NSF and DOE HEP.  Individual collaborators will propose to their respective funding agencies for scientific support to analyze the data including low level monitoring and calibration activities.
No decommissioning costs are expected because the recycling of the steel tanks can provide sufficient funds to cover returning the site to its original condition.

The total cost of 61.5M USD (54M USD for construction plus 7.5M USD for 5 years of operation) will be funded by an international collaboration as mentioned above. The exact cost sharing has yet to be determined, but it is anticipated that the US contribution would be about 20M USD.

\begin{table}
\begin{center}
\begin{tabular}{|l | r|}
\hline
\multicolumn{1}{|c |}{\textbf{Item}} & \multicolumn{1}{| c|}{\textbf{Cost (\$USD)}} \\
\hline
Structural detection unit (tank, bladder, roof, shipping) & 5k$\times$5000 = 25.0M\\
Photomultiplier-Tubes (8" and 3" per unit, encapsulation, base, high voltage) & 2k$\times$5000 = 10.0M\\
Electronics (read-out, trigger, timing, computing, network) & 5.0M\\
Water  (filtration system, purchase and delivery) & 3.0M\\
Infrastructure (site preparation, OPS building, DAQ building, AC power) & 5.0M\\
Labor for construction (100 person-years) & 3.0M \\
Management and Engineering  & 3.0M \\
\hline
\textbf{Total Construction Costs}  & \textbf{54M}\\
\hline
\end{tabular}
\end{center}
\caption{Cost breakdown.}\label{tab:cost}
\end{table}

\subsection*{\textcolor{BlueGreen}{Summary}}
\vspace{-.25cm}
In this white paper, we propose a next-generation gamma-ray observatory to be built in South America. SWGO with its wide field of view and continuous observations will provide the required sensitivity from below 1 TeV to beyond 100 TeV to constrain the extreme physics of the multi-messenger universe. The proven design and experience with HAWC ensure that SWGO will be built on time and on budget while delivering this required sensitivity. A strong international team of scientists is finalizing the design and site selection in order to make new discoveries with SWGO within the next decade.


\pagebreak

\section*{Affiliations}
\noindent
\footnotemark[1]{	Laboratório de Instrumentação e Física Experimental de Partículas (LIP), Portugal	}\\
\footnotemark[2]{	Instituto Superior Técnico (IST), Portugal	}\\
\footnotemark[3]{	Los Alamos National Lab, USA	}\\
\footnotemark[4]{	Instituto de Física, UNAM, Mexico	}\\
\footnotemark[5]{	Facultad de Ciencias en Física y Matemáticas-UNACH, Mexico	}\\
\footnotemark[6]{	Universidade de Lisboa, Portugal	}\\
\footnotemark[7]{	Pontificia Universidad Catolica del Peru, Peru	}\\
\footnotemark[8]{	The Ohio State University, USA	}\\
\footnotemark[9]{	The University of Adelaide, Australia	}\\
\footnotemark[10]{	University of Rochester, USA	}\\
\footnotemark[11]{	RWTH Aachen University, III. Physics Institute A, Germany	}\\
\footnotemark[12]{	Michigan Technological University, USA	}\\
\footnotemark[13]{	Durham University, UK	}\\
\footnotemark[14]{	IRFU, CEA Paris-Saclay, France }\\
\footnotemark[15]{	Università di Catania, INFN Catania, Italy 	}\\
\footnotemark[16]{	Universidad Autónoma de Chiapas, Mexico	}\\
\footnotemark[17]{	Università degli Studi di Roma "Tor Vergata", Roma, Italy	}\\
\footnotemark[18]{	Instituto Nacional de Astrofísica, Óptica y Electrónica, Tonantzintla, Mexico	}\\
\footnotemark[19]{	IFJ PAN, Poland	}\\
\footnotemark[20]{	Università agli Studi di Torino, Italy	}\\
\footnotemark[21]{	University of Oxford, UK	}\\
\footnotemark[22]{	Gran Sasso Science Institute, Italy	}\\
\footnotemark[23]{	Instituto de Astronomía y Física del Espacio (IAFE, CONICET-UBA), Argentina	}\\
\footnotemark[24]{	Departamento de Ciencias de la Atmósfera y los Océanos (DCAO, FCEN-UBA), Argentina	}\\
\footnotemark[25]{	University of Padova, Italy	}\\
\footnotemark[26]{	University of Udine, Italy	}\\
\footnotemark[27]{	INFN Padova, Italy	}\\
\footnotemark[28]{	Università Iuav Di Venezia, Italy	}\\
\footnotemark[29]{	Wisconsin IceCube Particle Astrophysics Center (WIPAC), University of Wisconsin -- Madison, USA	}\\
\footnotemark[30]{	INFN, Roma Tor Vergata, Italy	}\\
\footnotemark[31]{	Universidad de Guadalajara, Mexico	}\\
\footnotemark[32]{	Universidad Tecnica Federico Santa Maria, Chile	}\\
\footnotemark[33]{	Universität Würzburg, Germany	}\\
\footnotemark[34]{	University of Leicester, UK	}\\
\footnotemark[35]{	Department of Physics, University of Wisconsin -- Madison, USA	}\\
\footnotemark[36]{	Karlsruhe Institute of Technology, Germany	}\\
\footnotemark[37]{	Pennsylvania State University, USA	}\\
\footnotemark[38]{	Coimbra Polytechnic - ISEC, Coimbra, Portugal	}\\
\footnotemark[39]{	Instituto de Astronomía, UNAM, Mexico	}\\
\footnotemark[40]{	Erlangen Centre for Astroparticle Physics, University Erlangen-Nuernberg, Germany	}\\
\footnotemark[41]{	University of Maryland, College Park, USA	}\\
\footnotemark[42]{	University of Liverpool, UK	}\\
\footnotemark[43]{	Dip. di Fisica e Astronomia, Catania University 	}\\
\footnotemark[44]{	Max Planck Institute for Nuclear Physics, Heidelberg, Germany	}\\
\footnotemark[45]{	Institute for Cosmic Ray Research, the University of Tokyo, Japan	}\\
\footnotemark[46]{	Sorbonne Université, Université Paris Diderot, Sorbonne Paris Cité, CNRS/IN2P3, Laboratoire de Physique Nucléaire et de Hautes Energies, LPNHE, Paris, France	}\\
\footnotemark[47]{	Centro de Investigación en Computación. Instituto Politécnico Nacional, Mexico	}\\
\footnotemark[48]{	Instituto de Física de São Carlos, Universidade de São Paulo, São Carlos SP, Brasil	}\\
\footnotemark[49]{	Facultad de Ciencias Físico Matemáticas, Benemérita Universidad Autónoma de Puebla, Mexico	}\\
\footnotemark[50]{	Department of Particle Physics and Astrophysics, Weizmann Institute of Science, Israel	}\\
\footnotemark[51]{	Michigan State University, USA	}\\
\footnotemark[52]{	INAF, Italy	}\\
\footnotemark[53]{	Dipartimento Interateneo di Fisica dell'Università e del Politecnico di Bari, Italy	}\\
\footnotemark[54]{	Instituto de Astrofísica, Pontificia Universidad Católica de Chile, Chile	}\\
\footnotemark[55]{	University of Alabama, Tuscaloosa, USA	}\\
\footnotemark[56]{	DESY Zeuthen, Germany	}\\
\footnotemark[57]{	CBPF, Rio de Janeiro, Brazil	}\\
\footnotemark[58]{	University of Utah, USA	}\\
\footnotemark[59]{	Centro de Investigación y de Estudios Avanzados del IPN, Mexico	}\\

\pagebreak

\bibliographystyle{collab-etal}
\bibliography{references}

\end{document}